\documentclass{PoS}

\usepackage{amsmath}
\usepackage{commath}
\usepackage{graphicx}
\usepackage{amssymb}
\usepackage{subfigure}
\usepackage{bm}
\usepackage{slashed}
\usepackage{tikz}
\usepackage{braket}
\usepackage[labelsep=space]{caption}
\usepackage[capitalise]{cleveref}

\def\process{\mbox{ $B \to K^{*} \nu \bar \nu$ }}

\def\be{\begin{equation}}
\def\ee{\end{equation}}	
\def\bea{\begin{eqnarray}}
\def\eea{\end{eqnarray}}
\def\d{\mathrm{d}}
\def\beq{\begin{equation}}
\def\eeq{\end{equation}}

\def\nnb{\nonumber}

\def\rar{\rightarrow}
\def\nnb{\nonumber}

\def\ba{\begin{array}}
	\def\ea{\end{array}}

\title{Comparing AdS/QCD and Sum Rules predictions for $B\to K^*\nu \bar\nu$}

\ShortTitle{AdS/QCD vs SR predictions for $B\to K*\nu \bar\nu$}

\author{\speaker{Mohammad Ahmady}\thanks{This research is supported by NSERC Discovery Grants SAPIN-2017-00033 (M.A) and SAPIN-2017-00031 (R.S).}\\
        Department of Physics, Mount Allison University, Sackville, New Brunswick, Canada E4L 1E6\\
        E-mail: \email{mahmady@mta.ca}}

\author{Alexandre Leger\\
       Department of Physics, Mount Allison University, Sackville, New Brunswick, Canada E4L 1E6\\
        E-mail: \email{azleger@mta.ca}}
    
\author{Zoe McIntyre\\
	Department of Physics, Mount Allison University, Sackville, New Brunswick, Canada E4L 1E6\\
	E-mail: \email{zxmcintyre@mta.ca}}    
    
\author{Alexander Morrison\\
	Department of Physics, Mount Allison University, Sackville, New Brunswick, Canada E4L 1E6\\
	E-mail: \email{ahmorrison@mta.ca}} 

\author{Ruben Sandapen\\
	Department of Physics, Acadia University, Wolfville, Nova-Scotia, Canada B4P 2R6\\
	E-mail: \email{ruben.sandapen@acadiau.ca}}   

\abstract{Using the form factors obtained from holographic AdS/QCD and QCD sum rules, we predict the differential branching ratio and longitudinal polarization fraction for the rare $B\to K^* \nu \bar\nu$ decay.  This is an interesting decay channel as it does not suffer from hadronic uncertainties beyond the form factors.  We point out that the future measurement of the \process with around 30\% accuracy can discriminate between the two models.}

\FullConference{The International Conference on B-Physics at Frontier Machines - BEAUTY2018\\
		6-11 May, 2018\\
		La Biodola, Elba Island, Italy}

\begin{document}

\section{Introduction}

The flavor changing neutral current (FCNC) $b\to s$ is an excellent venue for searching signals of new physics (NP) beyond the Standard Model (SM).   The rare decay  \process is particularly interesting as, on the theory side, the presence of only one operator in the effective Hamiltonian for the $b\to s \nu\bar\nu$ transition implies sensitivity to a minimal number of form factors (FF).  Experimentally, however, is challenging, as both leptons are detector eluding neutrinos. The only available data at this time are the upper bounds on the branching ratio ($\mathcal{BR}$) set by the Belle Colaboration \cite{Lutz:2013ftz}:
\bea
\mathcal{BR}(B^+\to K^{*+}\nu\bar{\nu})&<& 4.0\times 10^{-5}\;\; (90\% \;\; {\rm CL})\;\; ,\nnb\\
\mathcal{BR}(B^0\to K^{*0}\nu\bar{\nu})&<& 5.5\times 10^{-5}\;\; (90\% \;\; {\rm CL})\;\; .
\label{upperbound}
\eea
However, the Belle-II experiment, with an integrated luminosity $50\; {\rm ab}^{-1}$ that
is expected to be collected by 2023, a measurement of the SM $\mathcal{BR}$s
with $30\%$ precision is expected \cite{Aushev:2010bq}. Therefore, it is appropriate to have a closer look at the theoretical calculation of this decay rate focusing on the theoretical uncertainties in the SM predictions.
Here, we present our results for the differential $\mathcal{BR}$ as well as the $K^*$ longitudinal polarization fraction for \process decay using the FFs parameterizing the $B\to K^*$ hadronic matrix elements which are derived via light-cone sum rules (LCSR).  The required Distribution Amplitudes (DAs) for $K^*$ are obtained from the holographic light-front wavefunctions (LFWFs) for vector mesons\cite{Ahmady:2014sva} and from QCD Sum Rules (SR) \cite{Straub:2015ica}.  The detailed calculations can be found in Ref \cite{Ahmady:2018fvo}.  We show that the two models make distinct predictions for $\mathcal{BR}$(\process ) specially at high $K^*$ recoil kinematic region.

%%%%%%%%%%%%%%%%%%%%%%%%%%%%%%%%%%%%%%%%%
\section{$K^*$ DAs and $B\to K^*$ FFs}
In terms of the LFWFs, the twist-$2$ DAs, $\phi_{K^*}^{\parallel ,\perp} (x,\; \mu )$, for $K^*$ are given as \cite{Ahmady:2013cva}:
\begin{equation}
f_{K^*} \phi_{K^*}^\parallel(x,\mu) =\sqrt{\frac{N_c}{\pi}} \int \d
b \mu
J_1(\mu b) \left[1 + \frac{m_{\bar{q}} m_{s} -\nabla_b^2}{M_{K^*}^2 x(1-x)}\right] \frac{\Psi_L(x, \zeta)}{x(1-x)} \;,
\label{phiparallel-phiL}
\end{equation}
and 
\begin{equation}
f_{K^*}^{\perp}(\mu) \phi_{K^*}^\perp(x,\mu) =\sqrt{\frac{N_c }{2 \pi}} \int \d
b \mu
J_1(\mu b) [m_s - x(m_s-m_{\bar{q}})] \frac{\Psi_T(x,\zeta)}{x(1-x)} \; ,
\label{phiperp-phiT}
\end{equation}
where $f_{K^*}$ and $f_{K^*}^\perp$ are the longitudinal and transverse coupling constants, respectively. $\mu\sim$ 1GeV is the nonperturbative hadronic scale and
\be  
\Psi_{\lambda} (x,\zeta) = \mathcal{N}_{\lambda} \sqrt{x (1-x)}  \exp{ \left[ -{ \kappa^2 \zeta^2  \over 2} \right] }
\exp{ \left[ -{(1-x)m_s^2+ xm_{\bar{q}}^2 \over 2 \kappa^2 x(1-x) } \right]} \;,
\label{hwf}
\ee
are holographic meson wavefunctions obtained by solving the holographic light-front Schr\"odinger Equation for mesons \cite{Brodsky:2014yha}. $\lambda =L,\; T$ denotes the polarization and $\zeta=\sqrt{x\bar x}b$ is the so-called holographic variable.  The longitudinal and transverse couplings are given by \cite{Ahmady:2013cva,Forshaw:2012im,Ahmady:2016ujw}
\bea
f_{K^*} &=&  {\sqrt \frac{N_c}{\pi} }  \int_0^1 {\mathrm d} x  \left[ 1 + { m_{\bar{q}}m_s-\nabla_{b}^{2} \over x (1-x) M^{2}_{K^*} } \right] \left. \Psi_L(\zeta, x) \right|_{\zeta=0}\; ,
\label{fvL}
\eea
and  
\begin{equation}
f_{K^*}^{\perp}(\mu) =\sqrt{\frac{N_c}{2\pi}}  \int_0^1 {\mathrm d} x (xm_{\bar{q}} +(1-x)m_s)  \int {\mathrm d} b  \; \mu J_1(\mu b)  \frac{\Psi_T(\zeta,x)}{x(1-x)} \; ,
\label{fvT}
\end{equation}
respectively.  As shown in Table \ref{tab:decay}, the quark masses can be constrained by fitting to both the measured decay constant $f_{K^*}$ and the lattice prediction for the ratio.  We observe that different sets of quark masses can be used to fit the measured decay constant with the larger quark masses being preferred in order to approach the lattice data for the ratio $f_{K^*}^{\perp}/f_{K^*}$. We shall use $m_{\bar{q}}=(195 \pm 55)$ MeV and $m_s=(300 \pm 20)$ MeV in computing our predictions. 

\begin{table}[h]
	%\begin{center}
	%\textbf{AdS/QCD predictions for the decay constants of $K^*$}
	\[
	\begin{array}
	[c]{|c|c|c|c|c|c|c|}\hline
	\mbox{Approach}&\mbox{Scale}~ \mu  &m_{\bar{q}} \mbox{[MeV]} & m_s \mbox{[MeV]} &f_{K^*} \mbox{[MeV]} &f_ {K^*}^{\perp} (\mu) \mbox{[MeV]}&f_{K^*}^{\perp}/f_{K^*} (\mu)\\ \hline
	%\rho & 0.548 & 0.140&0.140&0.214&0.135&0.63 \\ \hline
	\mbox{AdS/QCD} & \sim 1~\mbox{GeV} & 140 & 280 & 200  & 118 & 0.59 \\ \hline
	\mbox{AdS/QCD} & \sim 1~\mbox{GeV} & 195 & 300 & 200  & 132 & 0.66 \\ \hline
	\mbox{AdS/QCD} & \sim 1~\mbox{GeV} & 250 & 320 & 200  & 142 & 0.71 \\ \hline \mbox{Experiment}  & &  & &205\pm 6\footnote{From $\Gamma(\tau^- \to K^{*-} \nu_{\tau})$}  & & \\ \hline
	% \mbox{SR}  & 1 ~\mbox{GeV}& 220 \pm 5&  185 \pm 10& 0.82 \pm 0.06\\ \hline
	%\mbox{SR}  & 2 ~\mbox{GeV}& 220 \pm 5& 162 \pm 9 & 0.73 \pm 0.04\\ \hline
	\mbox{Lattice}  & 2 ~\mbox{GeV}& & & & &0.780 \pm 0.008 \\ \hline
	\mbox{Lattice}  & 2 ~\mbox{GeV}& &  & &  & 0.74 \pm 0.02\\ \hline
	\end{array}
	\]
	%\end{center}
	\caption {Comparison between AdS/QCD predictions for the decay constant of the $K^*$ meson with experiment \cite{Beringer:1900zz}, and the ratio of couplings with lattice \cite{Becirevic:2003pn,Braun:2003jg} data.}
	\label{tab:decay}
\end{table}

The twist-$2$ SR DAs are constructed as Gegenbauer expansions \cite{Ball:2007zt}
\begin{equation}
\phi_{K^*}^{\parallel,\perp}(x, \mu) = 6 x \bar x \left\{ 1 + \sum_{j=1}^{2}
a_j^{\parallel,\perp} (\mu) C_j^{3/2}(2x-1)\right\} \;. 
\label{phiperp-SR}
\end{equation}
where $C_j^{3/2}$ are the Gegenbauer polynomials and the coeffecients $a_j^{\parallel,\perp}(\mu)$ are given as $a_1^\parallel =0.06\pm 0.04$, $a_2^\parallel =0.16\pm 0.09$ for $\phi^\parallel(z,\; \mu=1\; {\rm GeV})$ and $a_1^\perp =0.04\pm 0.03$, $a_2^\perp =0.10\pm 0.08$ for $\phi^\perp(z,\; \mu=1\; {\rm GeV})$ \cite{Straub:2015ica}.

%%%%%%%%%%%%%%%%%%%%%%%%%%%%%%%%
\begin{figure}[htbp]
	\begin{subfigure}{}
		\centering
		\includegraphics[width=0.23\textwidth]{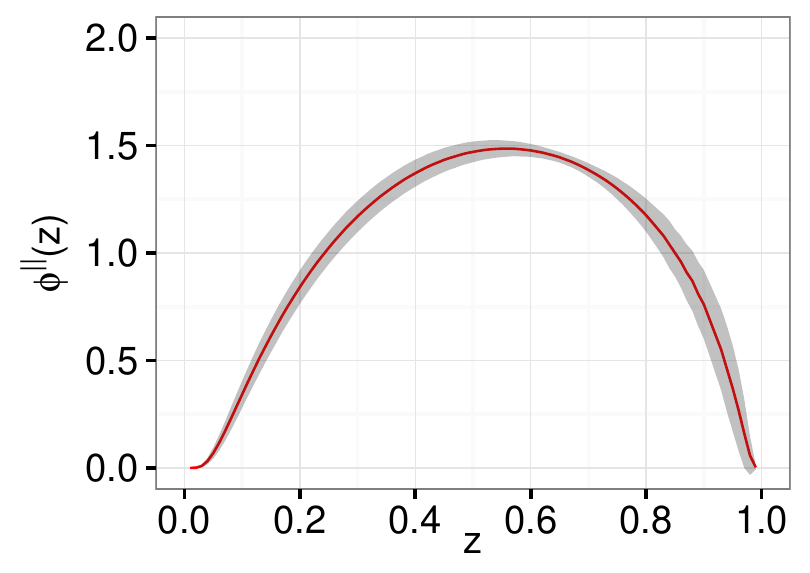}
		%\caption{1a}
		\label{adsdapara}
	\end{subfigure}
%	\hspace{.1cm}
	\begin{subfigure}{}
		\centering
		\includegraphics[width=0.23\textwidth]{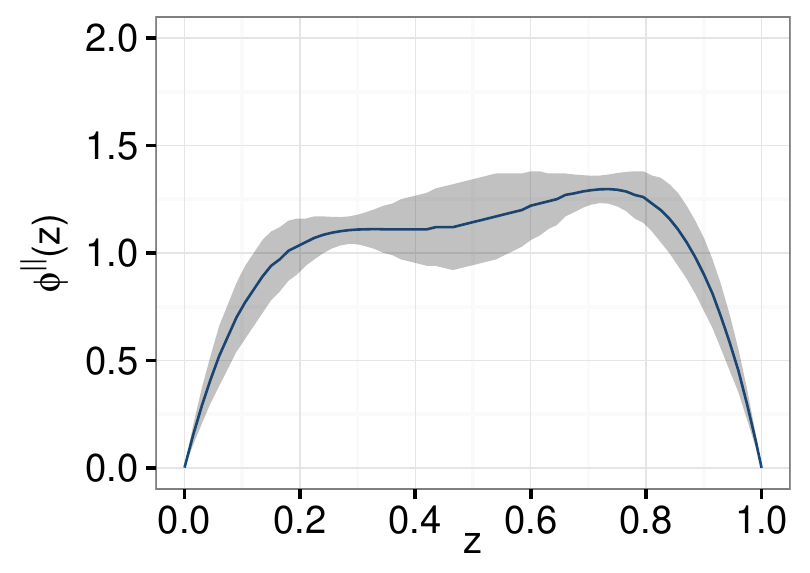}
		%\caption{1b}
		\label{srdapara}
	\end{subfigure}
	\begin{subfigure}{}
		\centering
		\includegraphics[width=0.23\textwidth]{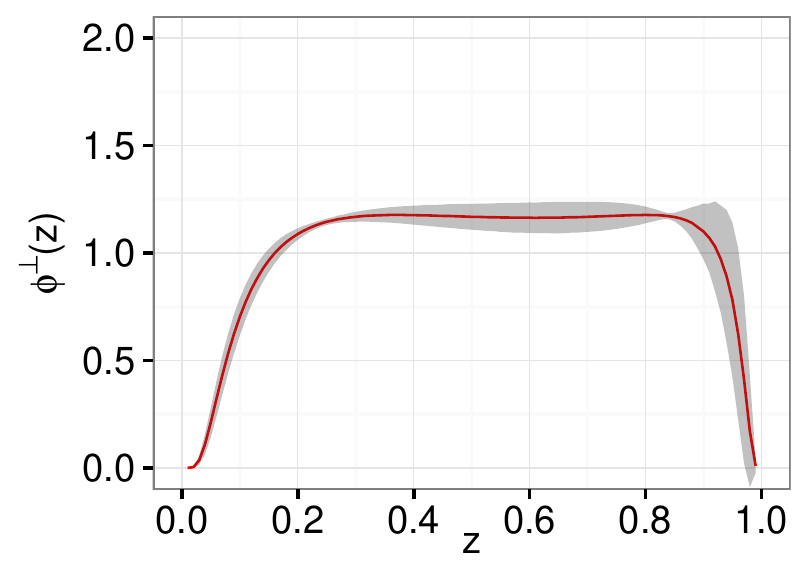}
		%\caption{1c}
		\label{adsdaperp}
	\end{subfigure}
	\begin{subfigure}{}
		\centering
		\includegraphics[width=0.23\textwidth]{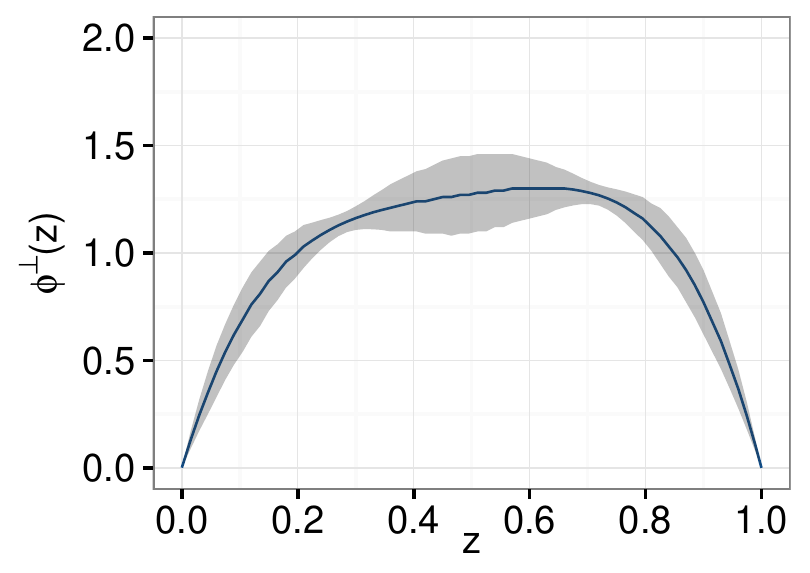}
		%\caption{1c}
		\label{srperp}
	\end{subfigure}
	\caption{Twist-2 DAs predicted by AdS/QCD (red curve graphs) and SR (blue curve graphs).}
	\label{das}
\end{figure}
Figure \ref{das} shows twist-2 DAs $\phi^{\parallel ,\perp}(z,\; \mu=1\; {\rm GeV})$ for the $K^*$ vector meson obtained from Eqs. \ref{phiparallel-phiL} and \ref{phiperp-phiT} as compared to SR predictions as given by Eq. \ref{phiperp-SR}.  The error band for holographic DAs are due to the uncertainty in the quark masses but the error band in SR DAs are the result of the uncertainties in the Gegenbauer coefficients. 

%%%%%%%%%%%%%%%%%%%%%%%%%%%%%%%%%%%%%%%%%
\section{$B\rar K^*$ transition FFs}
% % % % % % % % % % % % % % % % % % % % %
The FFs, computed via LCSR, are valid at low to intermediate $q^2$.  The extrapolation to high $q^2$ is performed via a two-parameter fit of the following form
%%%%%%%%%%%%%%%%%%%%%%%%%%%%%%%%%%%%%%%%%%
\begin{equation}
F(q^2)=\frac{F(0)}{1-a(q^2/m_B^2)+b(q^4/m_B^4)}
\end{equation}
to the LCSR predictions as well as form factor values obtained by the lattice QCD which are available at high $q^2$.

\begin{figure}[htbp]
	\begin{subfigure}{}
		\centering
		\includegraphics[width=0.3\textwidth]{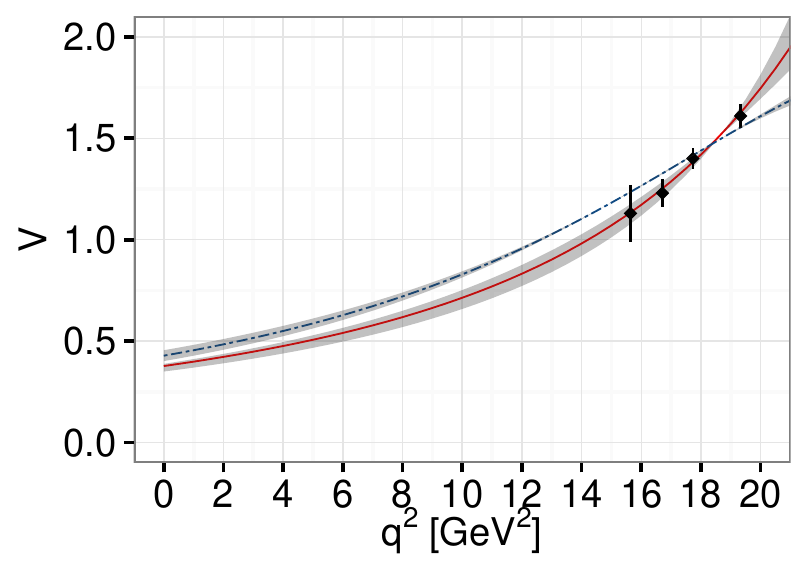}
		%\caption{1a}
		\label{fig:plot_V}
	\end{subfigure}
	\hspace{.1cm}
	\begin{subfigure}{}
		\centering
		\includegraphics[width=0.3\textwidth]{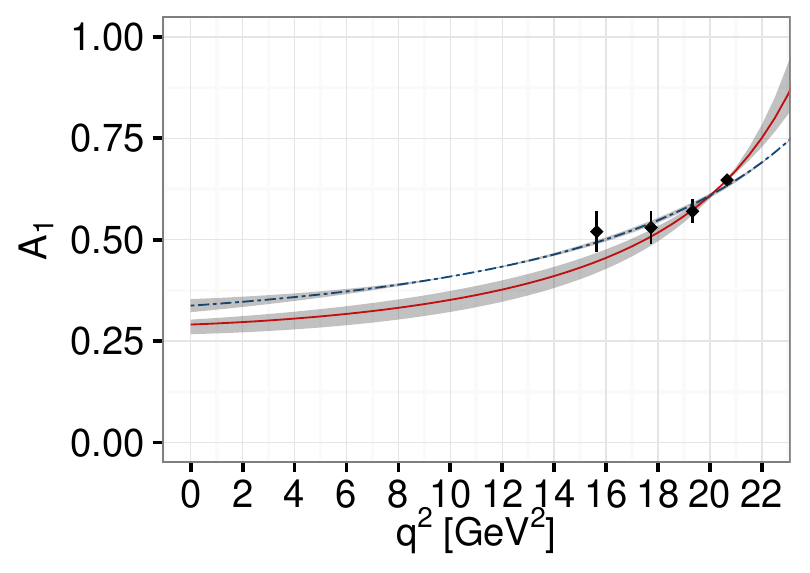}
		%\caption{1b}
		\label{fig:plot_A1}
	\end{subfigure}
	\begin{subfigure}{}
		\centering
		\includegraphics[width=0.3\textwidth]{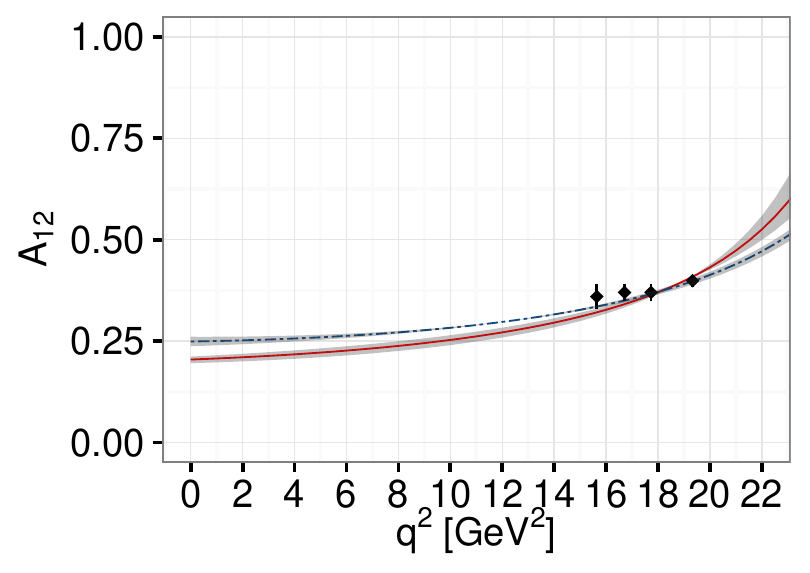}
		%\caption{1c}
		\label{fig:plot_A12}
	\end{subfigure}
	\caption{AdS/QCD predictions for the form factors $V$, $A_1$ and $A_{12}$. The two-parameter fits with the available lattice data (red) are shown and compared with the predictions of SR (dashed blue).}
	\label{formfactors}
\end{figure}

Figures \ref{formfactors} shows the AdS/QCD predictions including the lattice data points at high $q^2$ for the form factors $V$, $A_1$ and $A_{12}$.  The shaded bands in these figures represent the uncertainty due to the error band in the DAs.  Note that there is an additional uncertainly in the FFs inherent in the LCSR method (uncertainty in the Borel parameter, continuum threshold and other input parameters) which are the same in both models and are not included in our results.
%%%%%%%%%%%%%%%%%%%%%%%%%%%

\vspace{-0.7cm}
%%%%%%%%%%%%%%%%%%%%%%%%%%% 	
\section{Differential decay rate and $F_L$}
\begin{figure}[htbp]
	\begin{subfigure}{}
	\centering
	\includegraphics[width=0.5\textwidth]{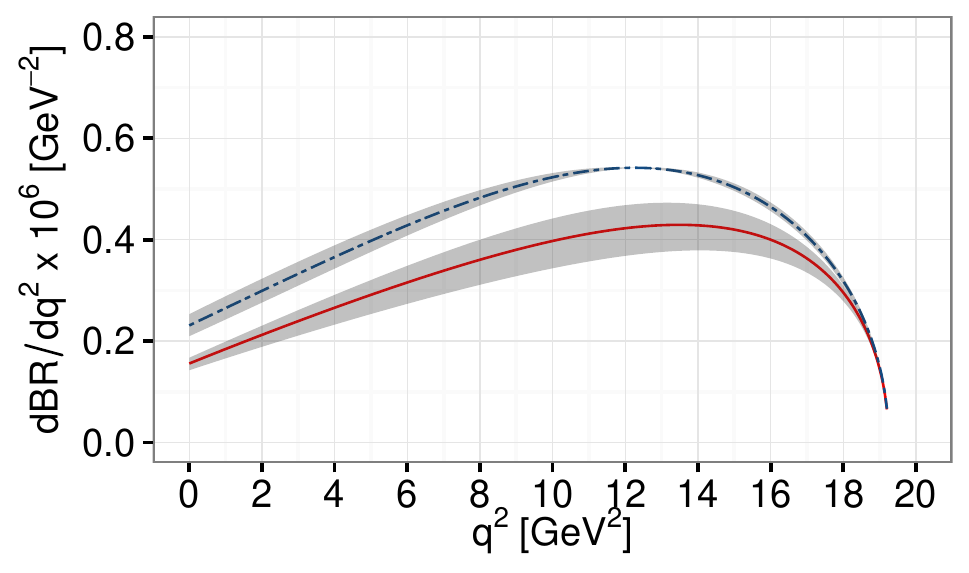}
	\end{subfigure}
	\hspace{.1cm}
	\begin{subfigure}{}
		\centering
		\includegraphics[width=0.5\textwidth]{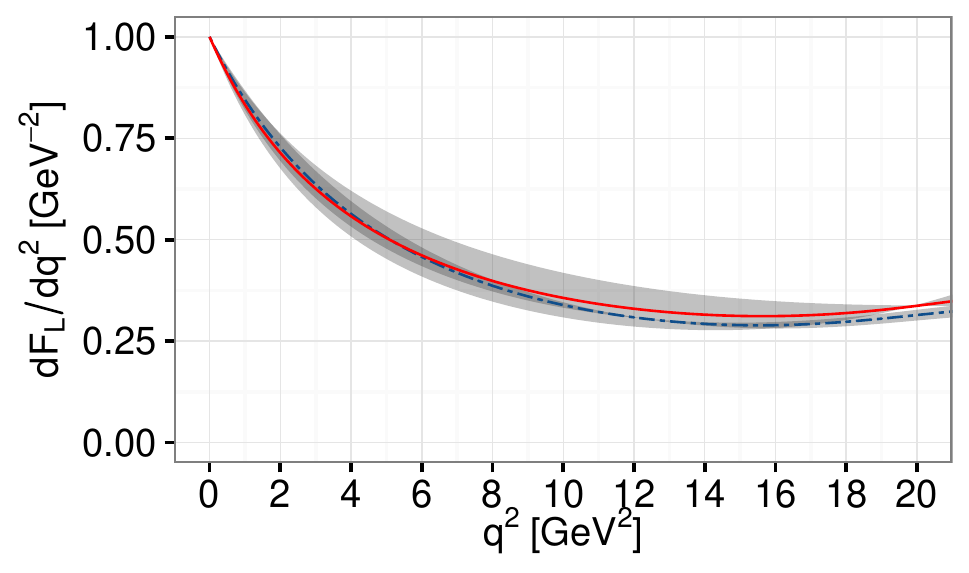}
	\end{subfigure}
	\caption{The AdS/QCD (Solid line) and SR (Dashed line) predictions for the differential Branching Ratio and polarization fraction $F_L$ for \process.}
	\label{fig:plot_BRKstar}
\end{figure}

%%%%%%%%%%%%%%%%%%%%%%%%%%%
 Figure \ref{fig:plot_BRKstar} compare the AdS/QCD and SR predictions for the \process differential decay rate and polarization fraction $F_L$.  The resulting uncertainties due to FFs are shown as the shaded bands.  We observe that, in general,  the AdS/QCD prediction is lower than SR prediction for all values of the momentum transfer $q^2$.  The difference between the two predictions is maximal ( $\sim 25\%$) for intermediate values of $q^2$.  Most interesting, the two predictions are quite distinct at low-to-intermediate $q^2$ where LCSR method is most reliable.    For the total branching ratio, we predict $\mathcal{BR}(\process)_{\rm AdS/QCD}=(6.36^{+0.59}_{-0.74})\times 10^{-6}$, compared to sum rules result $\mathcal{BR}(\process)_{\rm SR}=(8.14^{+0.16}_{- 0.17})\times 10^{-6}$.

  On the other hand, we observe that within error bands, the two model predictions for $F_L$ are not distinguishable.  This confirms that the $K^*$ longitudinal polarization fraction has little sensitivity to the non-perturbative FFs and is thus an excellent observable to probe New Physics signals.  Integrating over the whole kinematic region $0\le q^2\le (m_B-m_{K^*})^2$, we predict $F_L(\process)_{\rm SM}^{\rm AdS/QCD}=0.40^{+0.02}_{-0.01}$ as compared to $F_L(\process)_{\rm SM}^{\rm SR}=0.41\pm 0.01$.

%%%%%%%%%%%%%%%%%%%%%	
\section{Conclusion}%
%%%%%%%%%%%%%%%%%%%%%

AdS/QCD predicts lower \process decay rate than QCD sum rules.  We expect that a future measurement of this decay channel at BELLE II may be able to discriminate between the two models.

\bibliographystyle{JHEP}
\bibliography{BKstarnu}

\end{document}